\documentclass[twocolumn,a4,showpacs,preprintnumbers,amsmath,amssymb]{revtex4}
\usepackage{epsfig}

 \def\ket{\!>\,} \def\ack{\,|\,}
\def\156er{$^{156}$Er}
\def\158er{$^{158}$Er}
\def\160er{$^{160}$Er}
\def\162er{$^{162}$Er}
\def\164er{$^{164}$Er}
\def\168er{$^{168}$Er}
\def\170er{$^{170}$Er}
\begin{document}



\author{J. A. Sheikh$^{1}$, G. H. Bhat$^{1}$, Y. Sun$^{2,3}$, G. B.
Vakil$^{1}$, and R. Palit$^{4}$ }
\address{$^1$Department of Physics, University of Kashmir, Srinagar,
190 006, India \\
$^2$Department of Physics, Shanghai Jiao Tong University, Shanghai
200240, People's Republic of China \\
$^3$Joint Institute for Nuclear Astrophysics, University of
Notre Dame, Notre Dame, Indiana 46556, USA \\
$^4$Tata Institute of Fundamental Research, Colaba, Mumbai, 400 005,
India }

\title{Triaxial projected shell model study of $\gamma$-vibrational bands
in even-even Er isotopes}


\begin{abstract}

We expand the triaxial projected shell model basis to include
triaxially-deformed multi-quasiparticle states. This allows us to
study the yrast and $\gamma$-vibrational bands up to high spins for
both $\gamma$-soft and well-deformed nuclei. As the first
application, a systematic study of the high-spin states in
Er-isotopes is performed. The calculated yrast and $\gamma$-bands
are compared with the known experimental data, and it is shown that
the agreement between theory and experiment is quite satisfactory.
The calculation leads to predictions for bands based on one- and
two-$\gamma$ phonon where current data are still sparse. It is
observed that $\gamma$-bands for neutron-deficient isotopes of
$^{156}$Er and $^{158}$Er are close to the yrast band, and further
these bands are predicted to be nearly degenerate for high-spin
states.

\end{abstract}

\pacs{21.60.Cs, 21.10.Hw, 21.10.Ky, 27.50.+e}

\maketitle

\section{Introduction}

Recent experimental advances in nuclear spectroscopic techniques
following Coulomb excitations, in-elastic neutron scattering, and
thermal neutron capture have made it possible to carry out a
detailed investigation of $\gamma$-vibrational bands in atomic
nuclei \cite{FA88,Be96,BJ93}. These bands are observed in both
spherical and  as well as in deformed nuclei. In spherical nuclei,
the vibrational modes are well described using the harmonic phonon
model \cite{BM75,Rc20}. Although exact harmonic motion has never
been observed, there are numerous examples of nuclei exhibiting near
harmonic vibrational motion. As a matter of fact, one- and
two-phonon excitations have been reported in a large class of
spherical nuclei. In deformed nuclei, vibrational motion is possible
around the equilibrium of deformed shape configuration. The deformed
intrinsic shape is parameterized in terms of $\beta$ and $\gamma$
deformation variables. These parameters are related to the axial and
non-axial shapes of a deformed nucleus. The one-phonon vibrational
mode in deformed nuclei with no component of angular momentum along
the symmetry axis ($K=0$) is called $\beta$-vibration and the
vibrational mode with component of angular momentum along the
symmetry axis ($K=2$) is referred to as $\gamma$-vibration. The
rotational bands based on the $\gamma$-vibrational state are known
as $\gamma$-bands \cite{Gu95,WA93,BJ91}. One-phonon $\gamma$-bands
have been observed in numerous deformed nuclei in most of the
regions of the periodic table. There has also been reports on
observation of two-phonon $\gamma$-bands \cite{Fa96,Ha98}.

Several theoretical models have been proposed to study $\gamma$-
bands with varying degree of success. The quasiparticle phonon
nuclear model (QPNM) \cite{GY81,So92}, which restricts the basis to,
at the most, two phonon states, has led to the conclusion that
two-phonon collective vibrational excitations cannot exist in
deformed nuclei due to the Pauli blocking of important quasiparticle
components. On the other hand, the multi-phonon method (MPM)
\cite{LP88,JP88} embodies an entirely different truncation scheme.
It employs only a few collective phonons and restricts the basis to
all the corresponding multi-phonon states up to eight phonons. This
approach predicts that, for strongly collective vibrations, two
phonon $K^\pi = 4^+$ excitations should appear at an energy of about
2.6 times the energy of the one-phonon $K^\pi = 2^+$ state
\cite{BJ93,BS94}. On the other hand, the dynamic deformation model
(DDM) \cite{Ku84}, which is quite different from the models
mentioned above, constructs collective potential from a set of
deformed single-particle basis states accommodating eight major
oscillator shells. This model predicts a collective $K^\pi = 4^+$ at
almost 2 MeV.

All the above mentioned models (QPNM, MPM, and DDM) do not have
their wave functions as eigen-states of angular momentum. Strictly
speaking, these methods do not calculate the states of angular
momentum, but the $K$-states ($K$ is the projection of angular
momentum on the intrinsic symmetry axis). To apply these models, one
has to assume that $I\approx K$. However, since an intrinsic
$K$-state can generally have its components spread over the space of
angular momenta of $I\ge K$, the reliability of these approaches
depends critically on actual situation. As pointed out by Soloviev
\cite{So92}, it is quit desirable to recover the good
angular-momentum in the wave functions.

Some algebraic models including the extended version of the
interacting boson (sdg-IBM) \cite{AI75,YA86} and pseudo-symplectic
models \cite{CL87} have also been employed to study the $\gamma$-
excitation modes and predict high collectivity for the double
$\gamma$- vibration \cite{Ga96}.

Recently, the triaxial projected shell model (TPSM) has been
employed to describe $\gamma$- bands \cite{SH99,YS00}. This model
uses shell model diagonalization approach and in this sense, it is
similar to the conventional shell model approach except that the
basis states in the TPSM are triaxially deformed rather than
spherical. In the present version of the model, the intrinsic
deformed basis is constructed from the triaxial Nilsson potential.
The good angular momentum states are then obtained through exact
three-dimensional angular momentum projection technique. In the
final stage, the configuration mixing is performed by diagonalizing
the pairing plus quadrupole-quadrupole Hamiltonian in the projected
basis \cite{KY95,HS91}. The advantage of the TPSM is that it
describes the deformed single-particle states microscopically as in
QPNM, MPM, and DDM, but its total many-body states are exact eigen
states of angular momentum operator. Correlations beyond the
mean-field are introduced by mixing the projected configurations.

It is to be noted that an intrinsic triaxial state in the TPSM is a
rich superposition of different $K$-states. For instance, the
triaxial deformed vacuum state is composed of $K=0, 2, 4, \cdots$
configurations. The projected bands from these $K=0,2$ and 4
intrinsic states are the dominant components of the ground-,
$\gamma$-, and 2$\gamma$-bands, respectively \cite{YS00}.

In the earlier TPSM analysis for even-even nuclei, the shell model
space was very restrictive, including only 0-quasiparticle (qp)
state \cite{SH99,YS00,SP01,YS02,Bo02,YS07}. This strongly limited
the application of the TPSM to the low-spin and low-excitation
region only. It was not possible to study high-spin states because
multi-qp configurations will usually become important for states
with $I>10$ in the normally deformed rare-earth nuclei. In the
present work, the qp-space is enlarged to incorporate the
two-neutron-qp, two-proton-qp and four-qp configuration consisting
of two protons plus two neutrons. This large qp space is adequate to
describe the bands up to second bandcrossing \cite{HS91}. The
purpose of the present work is, as a first application of the
extended model, to perform a detailed investigation of the high-spin
band structures, in particular $\gamma$-bands, of Erbium isotopes
ranging from mass number $A=156$ to 170. In a parallel work
\cite{GC06}, the TPSM analysis for odd-odd nuclei in a multi-qp
space has been performed.

The manuscript is organized in the following manner: in the next
section, a brief description of the TPSM method is presented. The
results of the TPSM study are presented and discussed in section
III. Finally, the work is summarized in section IV.

\section{Triaxial Projected Shell Model approach}

In the present work, the TPSM qp basis is extended, which consists
of projected 0-qp vacuum, 2-proton (2p), 2-neutron (2n), and 4-qp
states, i.e.,
\begin{equation}
\begin{array}{r}
\hat P^I_{MK}\ack\Phi\ket ,\\
~~\hat P^I_{MK}~a^\dagger_{p_1} a^\dagger_{p_2} \ack\Phi\ket ,\\
~~\hat P^I_{MK}~a^\dagger_{n_1} a^\dagger_{n_2} \ack\Phi\ket ,\\
~~\hat P^I_{MK}~a^\dagger_{p_1} a^\dagger_{p_2} a^\dagger_{n_1}
a^\dagger_{n_2} \ack\Phi\ket .
\label{basis}
\end{array}
\end{equation}
In Eq. (\ref{basis}), the three-dimensional angular-momentum
operator is \cite{Ri80}
\begin{equation}
\hat P^I_{MK} = {2I+1 \over 8\pi^2} \int d\Omega\,
D^{I}_{MK}(\Omega)\, \hat R(\Omega),
\end{equation}
with the rotational operator
\begin{eqnarray}
\hat R(\Omega) = e^{-\imath \alpha \hat J_z} e^{-\imath \beta \hat J_y}
e^{-\imath \gamma \hat J_z},
\end{eqnarray}
and $\ack\Phi\ket$ represents the triaxial qp vacuum state. The qp
basis chosen above are adequate to describe the high-spin states
upto, say $I\sim 24$, and in the present analysis we shall restrict
to this spin regime. The triaxially deformed qp states are generated
by the Nilsson Hamiltonian
\begin{equation}
\hat H_N = \hat H_0 - {2 \over 3}\hbar\omega\left\{\epsilon\hat Q_0
+\epsilon'{{\hat Q_{+2}+\hat Q_{-2}}\over\sqrt{2}}\right\}.
\label{nilsson}
\end{equation}
Here $\hat H_0$ is the spherical single-particle Hamiltonian, which
contains a proper spin-orbit force \cite{Ni69}. The parameters
$\epsilon$ and $\epsilon'$ describe axial quadrupole and triaxial
deformations, respectively. It should be noted that for the case of
axial-symmetry, the qp vacuum state has $K=0$, whereas in the
present case of triaxial deformation, the vacuum state
$\ack\Phi\ket$ is a superposition of all the possible $K$-values.
The allowed values of the $K$-quantum number for a given intrinsic
state are obtained through the following symmetry consideration. For
the symmetry operator, $\hat S = e^{-\imath \pi \hat J_z}$, we have
\begin{equation}
\hat P^I_{MK}\ack\Phi\ket = \hat P^I_{MK} \hat S^{\dagger} \hat S
\ack\Phi\ket = e^{\imath \pi (K-\kappa)} \hat P^I_{MK}\ack\Phi\ket,
\end{equation}
where $\hat S\ack\Phi\ket = e^{-\imath \pi \kappa}\ack\Phi\ket$, and
$\kappa$ characterizes the intrinsic states in Eq. (\ref{basis}).
For the self-conjugate vacuum or 0-qp state, $\kappa=0$ and,
therefore, it follows from the above equation that only $K=$ even
values are permitted for this state. For 2-qp states, the possible
values for $K$-quantum number are both even and odd depending on the
structure of the qp state. For the 2-qp state formed from the
combination of the normal and the time-reversed states, $\kappa = 0$
and, therefore, only $K=$ even values are permitted. For the
combination of the two normal states, $\kappa=1$ and only $K=$ odd
states are permitted.

As in the earlier projected shell model (PSM) calculations, we use
the pairing plus quadrupole-quadrupole Hamiltonian \cite{KY95}
\begin{equation}
\hat H = \hat H_0 - {1 \over 2} \chi \sum_\mu \hat Q^\dagger_\mu
\hat Q^{}_\mu - G_M \hat P^\dagger \hat P - G_Q \sum_\mu \hat
P^\dagger_\mu\hat P^{}_\mu .
\label{hamham}
\end{equation}
The interaction strengths are taken as follows: The $QQ$-force
strength $\chi$ is adjusted such that the physical quadrupole
deformation $\epsilon$ is obtained as a result of the
self-consistent mean-field HFB calculation \cite{KY95}. The monopole
pairing strength $G_M$ is of the standard form
$$G_M = \left[21.24 \mp13.86(N-Z)/A\right]/A,$$
with ``$-$" for neutrons and ``$+$" for protons, which approximately
reproduces the observed odd--even mass differences in the rare-earth
mass region. This choice of $G_M$ is appropriate for the
single-particle space employed in the PSM, where three major shells
are used for each type of nucleons ($N=4,5,6$ for neutrons and
$N=3,4,5$ for protons). The quadrupole pairing strength $G_Q$ is
assumed to be proportional to $G_M$, and the proportionality
constant being fixed as 0.18. These interaction strengths are
consistent with those used earlier for the same mass region
\cite{SH99,KY95,YS00}.

\section{Results and Discussions}

The triaxial projected shell model calculations have been performed
for Er-isotopes ranging from $A$ = 156 to 170. The deformation
parameters $(\epsilon, \epsilon')$ used in the present work are same
as those employed in Ref. \cite{YS00}. It has already been mentioned
in section II that in the present work the mean-field potential is
constructed with given input deformation values of $\epsilon$ and
$\epsilon'$. In a more realistic calculation, these deformation
values for a given system are obtained through the variational HFB
calculations. The chosen values of $\epsilon$ for the present
calculation are those from the measured quadrupole deformations of
the nuclei as is done in the previous projected shell model
analysis. The $\epsilon'$ values used in the present work are
realistic, which correctly reproduce, for example, excitations of
the $\gamma$ band relative to the ground-state \cite{YS00}.

\subsection{Band Diagrams}

Band diagrams can bring valuable information regarding the
underlying physics \cite{KY95}. These band diagrams for the studied
Er-isotopes are presented in Figs. 1 to 4 and depict the results of
the projected energies for each intrinsic configuration. In the
diagrams, the projected energies are shown for 0, 2n, 2p and 2p+2n
quasiparticle configurations. The qp energies for these
configurations are given in the legend of each figure. As already
mentioned in the last section that with the triaxial basis, the
intrinsic states do not have a well-defined $K$-quantum number. Each
triaxial configuration in Eq.(\ref{basis}) is a composition of
several $K$-values and bands in Figs. 1 to 4 are obtained by
assigning a given $K$-value in the angular-momentum projection
oprator. To make the discussion easy, we denote a $K$-state of an
$i$-configuration as $(K,i)$, with $i=0$, 2n, 2p and 4. For example,
the $K=0$ state of 0-qp configuration is marked as $(0,0)$ and $K=1$
of 2n-qp configuration as (1, 2n).

In Figs. 1 to 4, the projected bands associated with the 0-qp
configuration are shown for $K=0, 2$, and 4, namely the (0, 0), (2,
0), and (4, 0) bands. In the literature,these $K=0,2$, and 4 bands
are referred to as ground-state, $\gamma$-, and $2\gamma$-bands. The
ground-state band has $\kappa=0$ and is, therefore, comprised of
only even-$K$ values. We use the same names in the following
discussion to be consistent with the literature, but stress that in
our final results obtained after diagonalisation, $K$ is not a
strictly conserved quantum number due to configuration mixing.

It is evident from Fig. 1 that the (2, 0) bands for $^{156}$Er and
$^{158}$Er lie very close to the (0, 0) bands. This means that
$\gamma$- vibration has low excitation energy in these two nuclei.
For high-spin states, it is further noted that the (0, 0) and (2, 0)
band energies become almost degenerate, and as a matter fact for
$I=16$ and above, the energy of even-spin states in the (2, 0) band
is slightly lower than the (0, 0) band. It is a well-known fact that
$\gamma$-bands become lower in energy with increasing triaxility and
what is also evident from Fig. 1 that they become favored with
increasing angular-momentum. As can be seen from Fig. 1, the (2, 0)
bands in $^{156}$Er and $^{158}$Er also depict pronounced signature
splitting with the splitting amplitude increasing with spin. The (4,
0) band is close to the (2, 0) band for $^{156}$Er and lies at a
slightly higher excitation energy for $^{158}$Er. The (4, 0) bands
in these two isotopes are also noted to have signature splitting for
higher angular momenta, and the splitting amplitude is nearly the
same for the (2, 0) and (4, 0) bands.

In Fig. 1, several representative multi-qp bands, namely projected
2- and 4-qp configurations, are also plotted. Although the $K=1$
2-qp neutron (1, 2n) and 2-qp proton (1, 2p) bands are close in
energy for low spins, but with increasing spin the 2n-qp bands are
lower in energy than 2p-qp bands due to larger rotational alignment.
It is noted that neutrons are occupying 1i$_{13/2}$ and protons are
occupying 1h$_{11/2}$ intruder sub-shells. For each of the (1, 2n)
and (1, 2p) bands, the projected energies are also shown for the
corresponding $\gamma$-bands with configurations (3, 2n) and (3,
2p). The (1, 2n) band  is noted to cross the (2, 0) and the (0, 0)
bands at $I=12$. It is also seen that the (3, 2n) band crosses the
(0, 0) band at a slightly higher spin value of $I=14$. It is
interesting to note that after the band crossing, the lowest
even-spin states originate from the (1, 2n) band, whereas the
odd-spin members are the projected states from the (3, 2n)
configuration. Finally, the 4-qp (4, 4) configuration lies at high
excitation energies and does not become yrast, at-least, up to the
spin values shown in the figure.

The band diagrams for $^{160}$Er and $^{162}$Er are presented in
Fig.2. The energy separation between the (0, 0) and (2, 0) bands is
larger as compared to the two lighter isotopes in Fig. 1. In the
case of $^{160}$Er, the (2, 0) band energies do come close to the
(0, 0) energies for spins $I > 12$. The (1, 2n) band again crosses
the (0, 0) band at $I=12$ for $^{160}$Er and at $I=14$ for
$^{162}$Er. The band diagrams for $^{164}$Er and $^{166}$Er shown in
Fig. 3 depict larger energy gaps among various bands. The signature
splitting of the (2, 0) band has considerably reduced. It is further
noted that 2n-band-crossing is shifted to higher spin values. For
the case of $^{164}$Er, the band crossing is observed to occur at
$I=16$ and for $^{166}$Er it occurs at $I=18$. The band diagrams for
$^{168}$Er and $^{170}$Er shown in Fig. 4 indicate that the (2, 0)
bands are quite high in excitation energy. The band crossing for
these cases is further shifted to higher spin values.

\subsection{Results after Configuration Mixing}

In the second stage of the calculation, the projected states
obtained above are employed to diagonalize the shell model
Hamiltonian of Eq. (\ref{hamham}). It is to be mentioned that for
the discussion purpose, only the lowest three bands from the 0-qp
configuration and lowest two bands for each other configuration have
been shown in band diagram, Figs. 1 to 4. However, in the
diagonalisation of the Hamiltonian, the basis states employed are
much more, which includes, for example, those $K=1,3,5$ and 7 with
$\kappa=1$ and $K=0,2,4,6$ and 8 with $\kappa=0$.

The lowest three bands after the configuration mixing are shown in
Figs. 5 and 6 and are compared with the experimental energies
wherever available. Although they are of mixed configurations in our
model, we still call them yrast, $\gamma$- and 2$\gamma$-bands to be
consistent with the literature. It is observed from these two
figures that the agreement between the calculated and the
experimental energies for the yrast and $\gamma$-bands is quite
satisfactory. For $^{156-164}$Er, the theoretical yrast line depicts
two slopes and these correspond to the slopes of two crossing bands
shown in Figs. 1 and 4. This also indicates that the interaction
between the two crossing bands is small with the result that these
nuclei shall depict a back-bending effect \cite{HS91}. It is also
encouraging to note from Figs. 5 and 6 that the agreement for the
$\gamma$-bands is quite good, except that for $^{164}$Er and
$^{170}$Er, the signature splitting at the top of the bands is not
reproduced properly. For the 2$\gamma$-bands, our calculations agree
well with the only available data in $^{166}$Er \cite{Fa96} and
$^{168}$Er \cite{Ha98}.

There is another notable effect about anharmonicity in $\gamma$
vibrations. If we regard the $\gamma$-bandhead as one $\gamma$-
phonon vibration and the 2$\gamma$-bandhead as two $\gamma$-phonon
vibration, it can be easily seen from Figs. 5 and 6 that the
vibration is not perfectly harmonic. In fact, in the two lightest
isotopes, the $\gamma$-soft $^{156}$Er and $^{158}$Er, the vibration
is almost harmonic. As the neutron number increases, a clear
anharmonicity is predicted from our calculation and the degree of
anharmonicity increases with increasing neutron number.

\subsection{Analysis of Wavefunction}

In order to probe further the structure of the bands presented in
Figs. 5 and 6, the wavefunction decomposition of the yrast,
$\gamma$- and 2$\gamma$-bands are shown in Figs. 7, 8 and 9 for
$^{156}$Er, $^{164}$Er and $^{170}$Er. For other nuclei, the
wavefunction have smiliar structure and are not presented. It is
seen from Fig. 7 that the yrast band for $^{156}$Er is predominantly
composed of the (0, 0) configuration up to $I=10$. The (0, 0)
contribution suddenly drops at $I=10$, and (1, 2n) configuration
becomes dominant from $I=12$ to 16. For $I=18$ and onwards, there
are many configurations with finite values contributing to the yrast
states. The band diagram of $^{156}$Er in Fig. 1 suggests that the
$\gamma$-band should have the (2, 0) configuration as the dominant
component. This is evident from Fig. 7 and it is also noted that (0,
0) is significant for the even spin states up to $I=8$. The $I=10$
state is mostly composed of (1, 2n) and for higher spin states the
(3, 2n) and (3, 2p) configurations are the dominant components of
the $\gamma$ band. The 2$\gamma$ band in Fig. 7 is composed of (4,
0) band for the low spin states. $I=8$ of this band is predominantly
composed of the (1, 2n) configuration, but the high spin states are
found to have quite a complex structure.

The yrast wavefunction decomposition of $^{164}$Er, shown in the top
panel of Fig. 8, indicates that this lowest band is predominantly
composed of the (0, 0) configuration up to $I=12$ and there appears
to be very small admixtures of $K=2$ and other configurations. After
the bandcrossing at $I=16$, the yrast states are dominated by the
(1, 2n) configuration. There is also a significant contribution of
the (3, 2n) configuration after the bandcrossing. The $\gamma$-band
in Fig. 8 is primarily composed of the (2, 0) configuration up to
$I=11$ and above this spin the states are a mixture of different
configurations. There is a clear distinction in the composition of
the even- and odd-spin states above $I=11$. The odd-spin states are
composed of the (3, 2n) and (2, 0) configurations, and the even-spin
states are dominated by the (1, 2n) and (0, 0) structures. The
2$\gamma$-band up to $I=7$ is primarily the (4, 0) configuration.
For $I=8$ and above, this band is a mixture of (1, 2n) and (3, 2n)
configurations.

The wavefunction analysis of $^{170}$Er shown in Fig. 9 indicates
that the yrast state, as expected for a well deformed nuclei, is
mainly comprised of the (0, 0) configuration. This contribution
drops smoothly and, on the other hand, the (1, 2n) component
increases steadily. For $I=20$, it is noted that the (0, 0) and (1,
2n) contributions are almost identical and above this spin value, it
is expected that the (1, 2n) configuration shall dominate the yrast
states. The $\gamma$-band is also noted to have a well defined
structure of (2, 0) and only for high spin states, it is observed
that the (1, 2n) and (3, 2n) of the 2n-aligned configuration become
important. The 2$\gamma$-band is dominated mostly by the aligning
configurations above $I=7$. As is evident from the band diagram of
this nucleus,presented in Fig. 4, that 2n-aligned band is lower than
the (4, 0) band for most of the spin values.

\section{Summary and Conclusions}

In the present work, the triaxial projected shell model approach
with extended basis has been employed to study the high-spin band
structures of the Er-isotopes from $A=156$ to 170. In this model,
the Hamiltonian employed consists of pairing plus
quadrupole-quadrupole interaction. It is known that Nilsson deformed
potential is the mean-field of the quadrupole-quadrupole interaction
and this potential is directly used as the Hartree-Fock field rather
than performing the variational calculations. It is, as a matter of
fact, quite appropriate to use the Nilsson states as a starting
basis because the parameters of this potential have been fitted to
large body of experimental data. The parameters of the model are the
deformation parameters of $\epsilon$ and $\epsilon'$. The axial
deformation parameter $\epsilon$ has been fixed from the observed
quadrupole deformation of the system as is done in most of the
projected shell model analysis. The non-axial parameter $\epsilon'$
was chosen to reproduce the bandhead of the $\gamma$ band. The
pairing strength parameters have been determined to reproduce the
odd-even mass differences.  The monopole pairing interaction has
been solved in the BCS approximation and the qp states generated. In
the present work, the qp states considered are: 0-qp, 2-qp neutron,
2-qp proton, and the 4-qp state of 2-neutron plus 2-proton.

In the second stage of the calculations, the three-dimensional
angular-momentum projection is performed to project out the good
angular-momentum states from these qp states. These projected states
are then used as the basis to diagonalise the shell model
Hamiltonian in the third and the final stage. The salient features
of results obtained in the present work are:

\begin{enumerate}

\item  $\gamma$-bands are quite close to the yrast line for the
neutron-deficient Er-isotopes, in particular, for $^{156}$Er and
$^{158}$Er. It is further evident from the present results that
these $\gamma$-states become further lower in energies for high-spin
states. As a matter of fact, for $^{156}$Er and $^{158}$Er, they
become lower than the ground-state band for $I>14$. We propose that
this is a feature of $\gamma$-soft nuclei.

\item $\gamma$-bands are pushed up in energy with increasing
neutron number, and further the degree of anharmonicity of $\gamma$
vibration also increases.

\item The wavefunction decomposition of the bands demonstrates
that for neutron deficient Er-isotopes, there is a significant mixture
of the $\gamma$ configuration in the ground-state band and
vice-versa. The neutron rich $^{170}$Er nucleus, on the other hand,
has the intrinsic structures as expected for a well deformed nucleus
with the ground-state band composed of nearly pure $K=0$
configuration.

\end{enumerate}

Y.S. is supported by the Chinese Major State Basic Research
Development Program through grant 2007CB815005, and by the the U. S.
National Science Foundation through grant PHY-0216783.


\newpage

\begin{figure*}
\begin{center}
\includegraphics[width=6.5in]{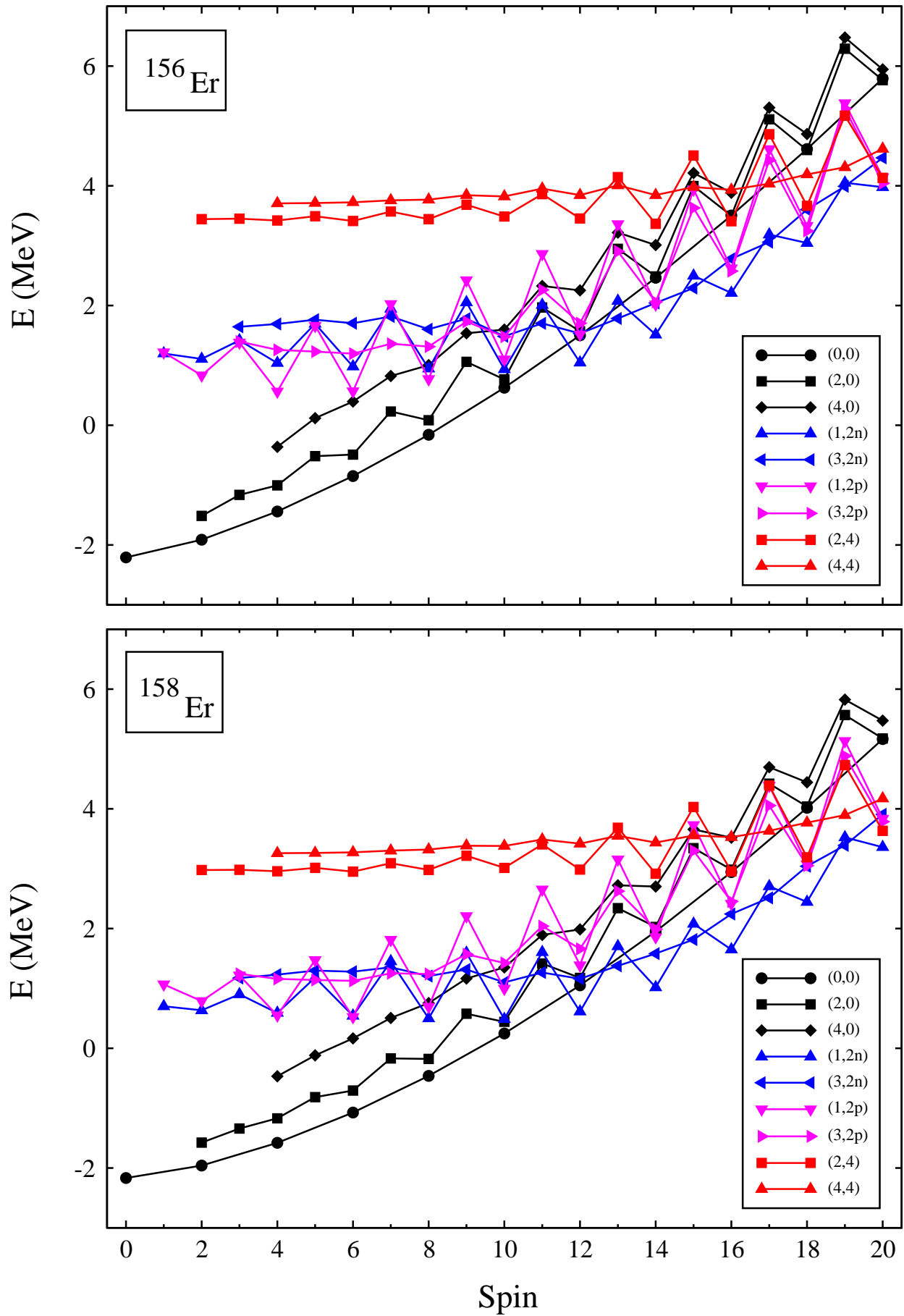}
\caption{
Band diagrams for $^{156-158}$Er isotopes. The labels (0,0), (2,0), (4,0),
(1,2n), (3,2n), (1,2p), (3,2p), (2,4) and (4,4) correspond to ground,
$\gamma$, 2$\gamma$, two neutron-aligned, $\gamma$-band on
this two neutron-aligned state, two proton-aligned, $\gamma$-band on
two this proton-aligned state, two-neutron plus two-proton aligned band and
$\gamma$ band built on this four-quasiparticle state.
}

\label{figure.1}
\end{center}
\end{figure*}

\begin{figure*}
\begin{center}
\includegraphics[width=6.5in]{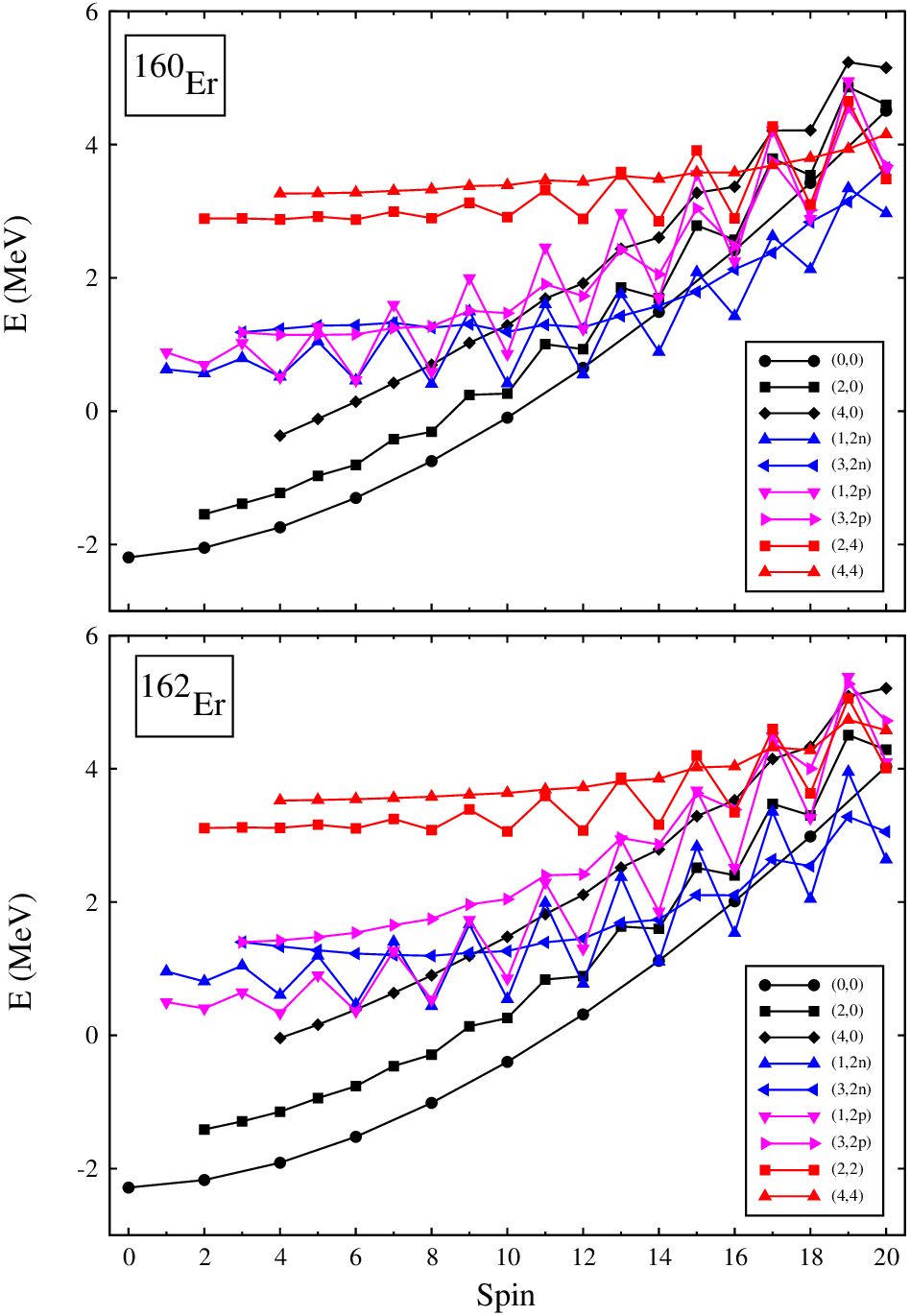}
\caption{
Band diagrams for $^{160-162}$Er isotopes. The labels indicate the bands
mentioned in the caption of Fig. 1.
}

\label{figure.2}
\end{center}
\end{figure*}

\begin{figure*}
\begin{center}
\includegraphics[width=6.5in]{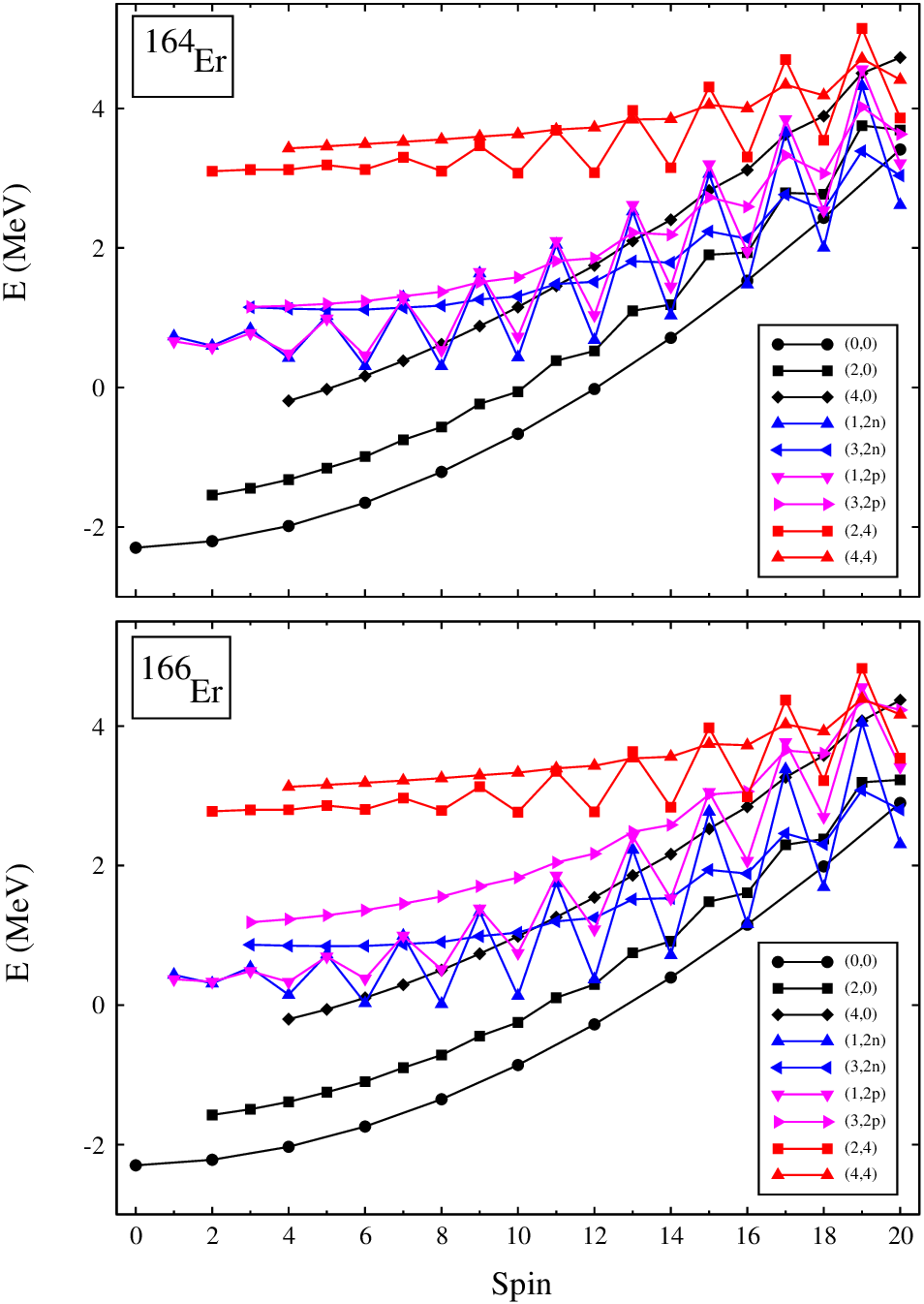}
\caption{
Band diagrams for $^{164-166}$Er isotopes. The labels indicate the bands
mentioned in the caption of Fig. 1.
}

\label{figure.3}
\end{center}
\end{figure*}

\begin{figure*}
\begin{center}
\includegraphics[width=6.5in]{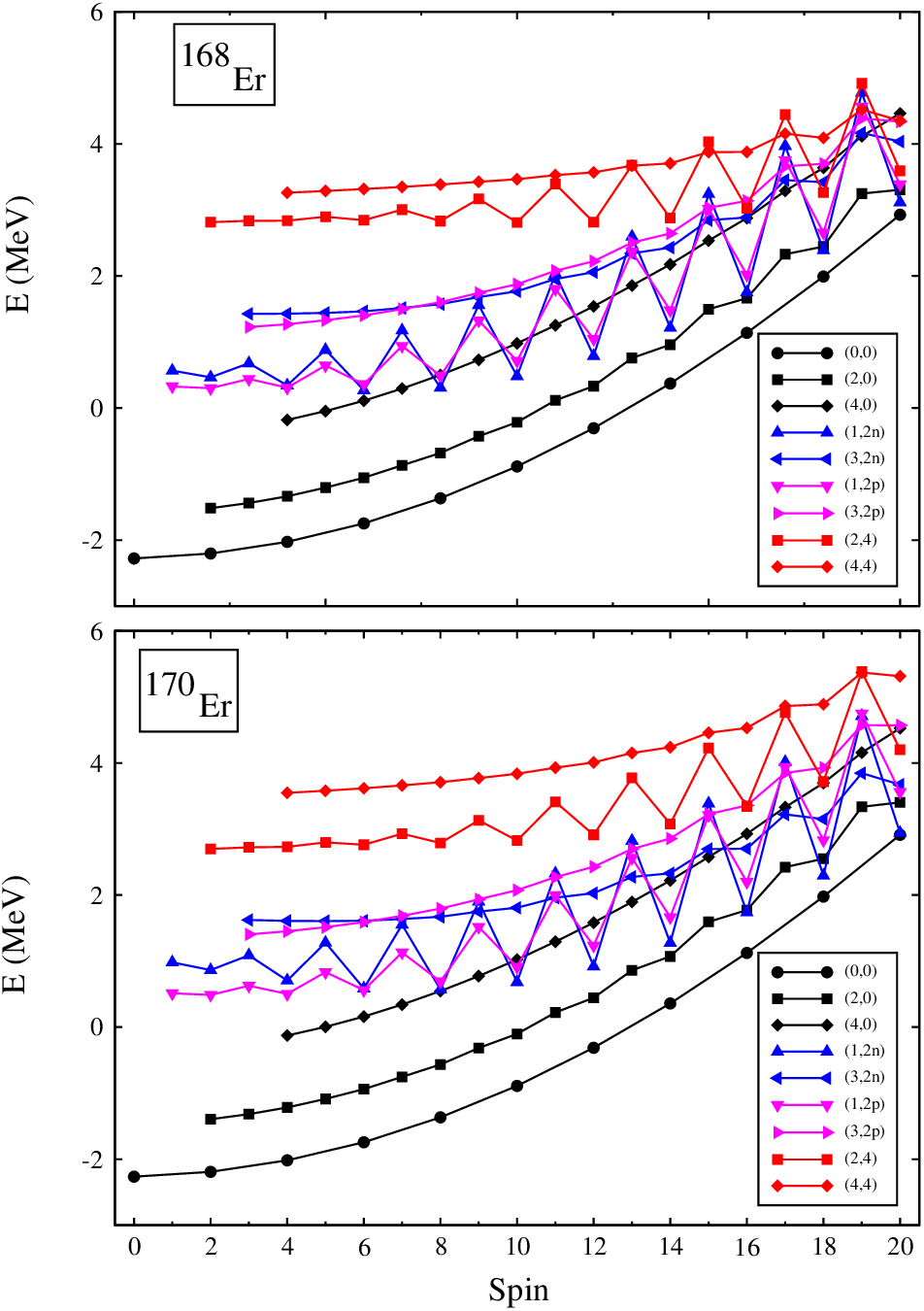}
\caption{
Band diagrams for $^{168-170}$Er isotopes. The labels indicate the bands
mentioned in the caption of Fig. 1.
}

\label{figure.4}
\end{center}
\end{figure*}

\begin{figure*}
\begin{center}
\includegraphics[width=6.5in]{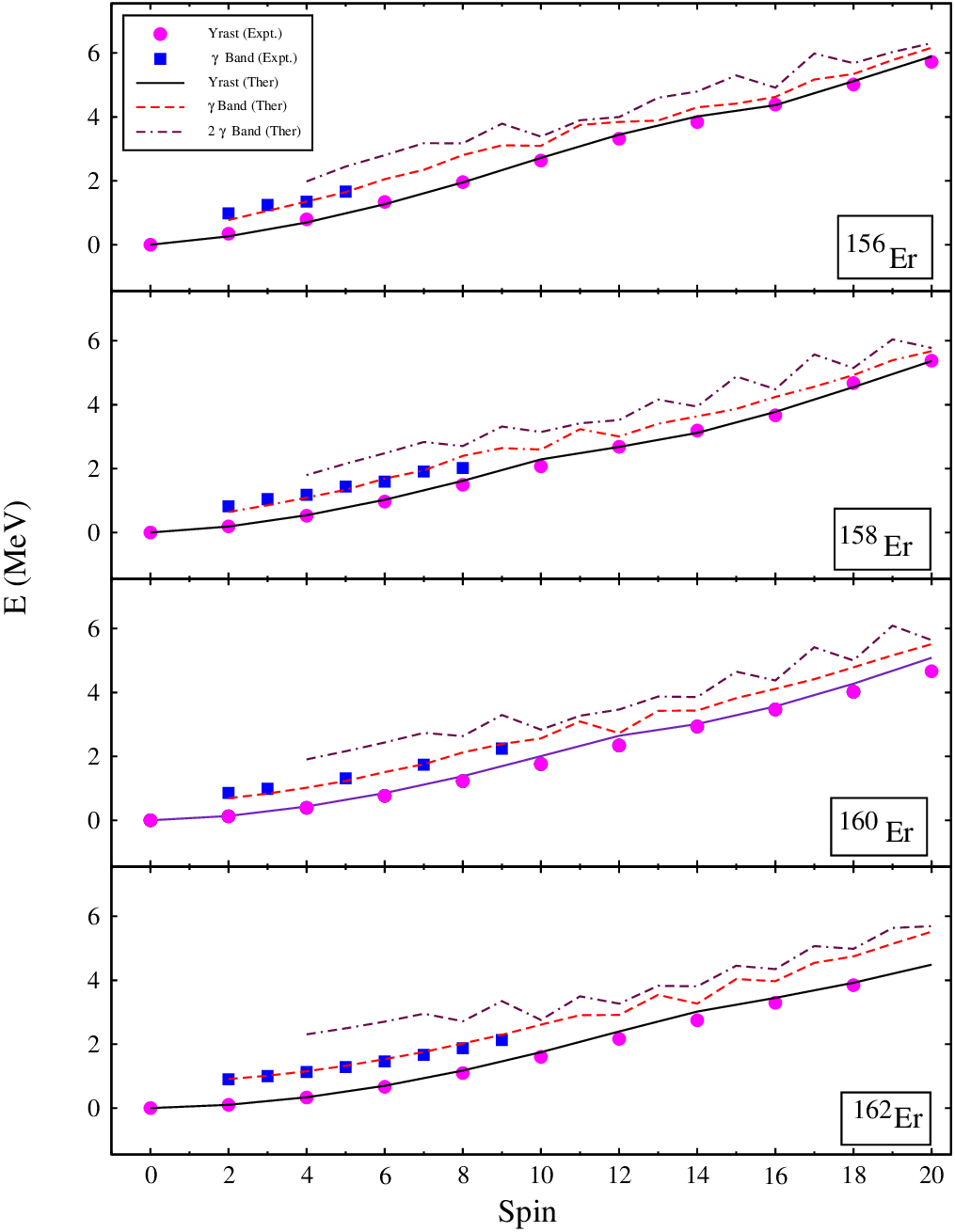}
\caption{ Comparison of experimental and the calculated band
energies for $^{156-162}$Er. }

\label{figure.5}
\end{center}
\end{figure*}

\begin{figure*}
\begin{center}
\includegraphics[width=6.5in]{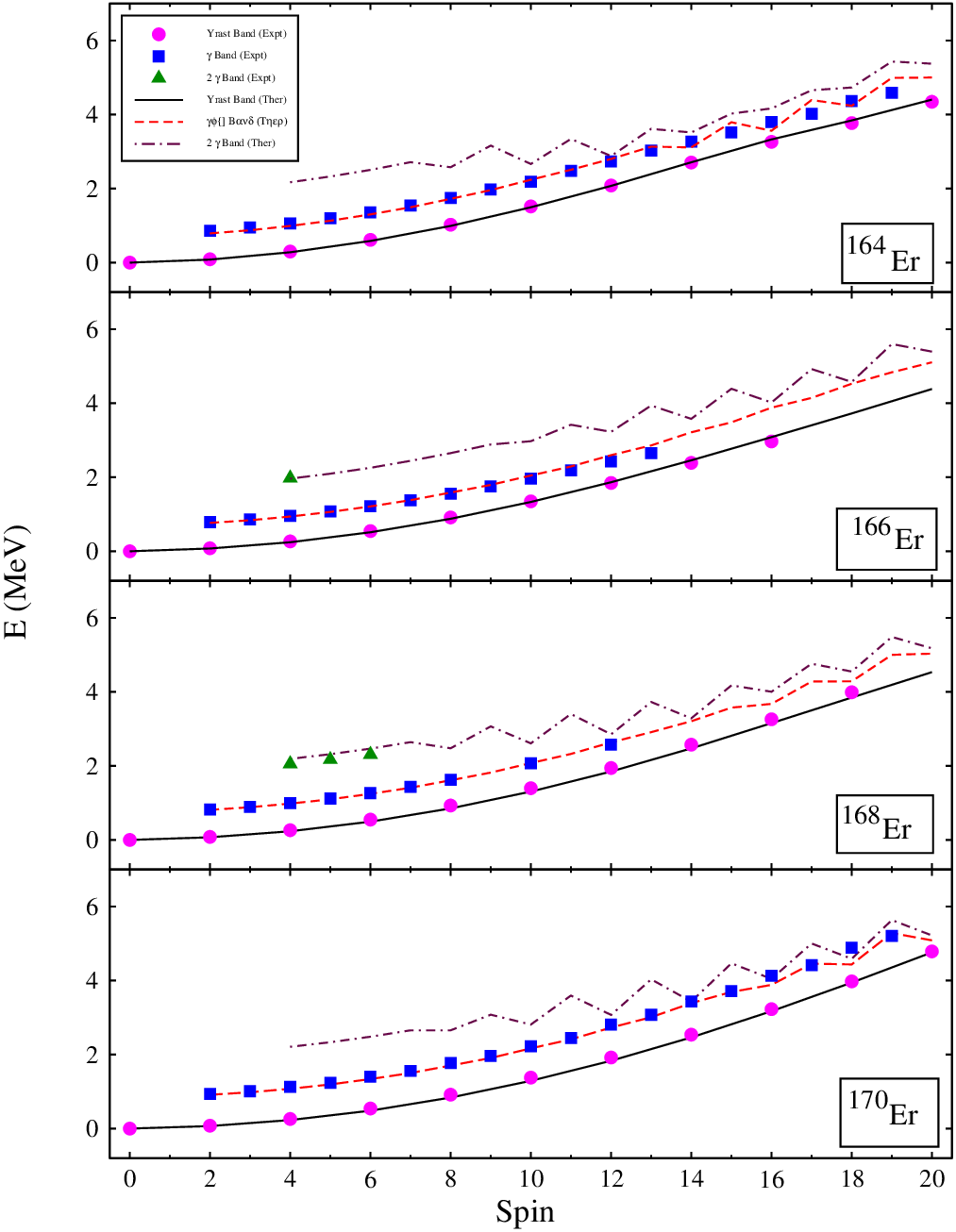}
\caption{ Comparison of experimental and the calculated band
energies for $^{164-170}$Er. }

\label{figure.6}
\end{center}
\end{figure*}

\begin{figure*}
\begin{center}
\includegraphics[width=6.5in]{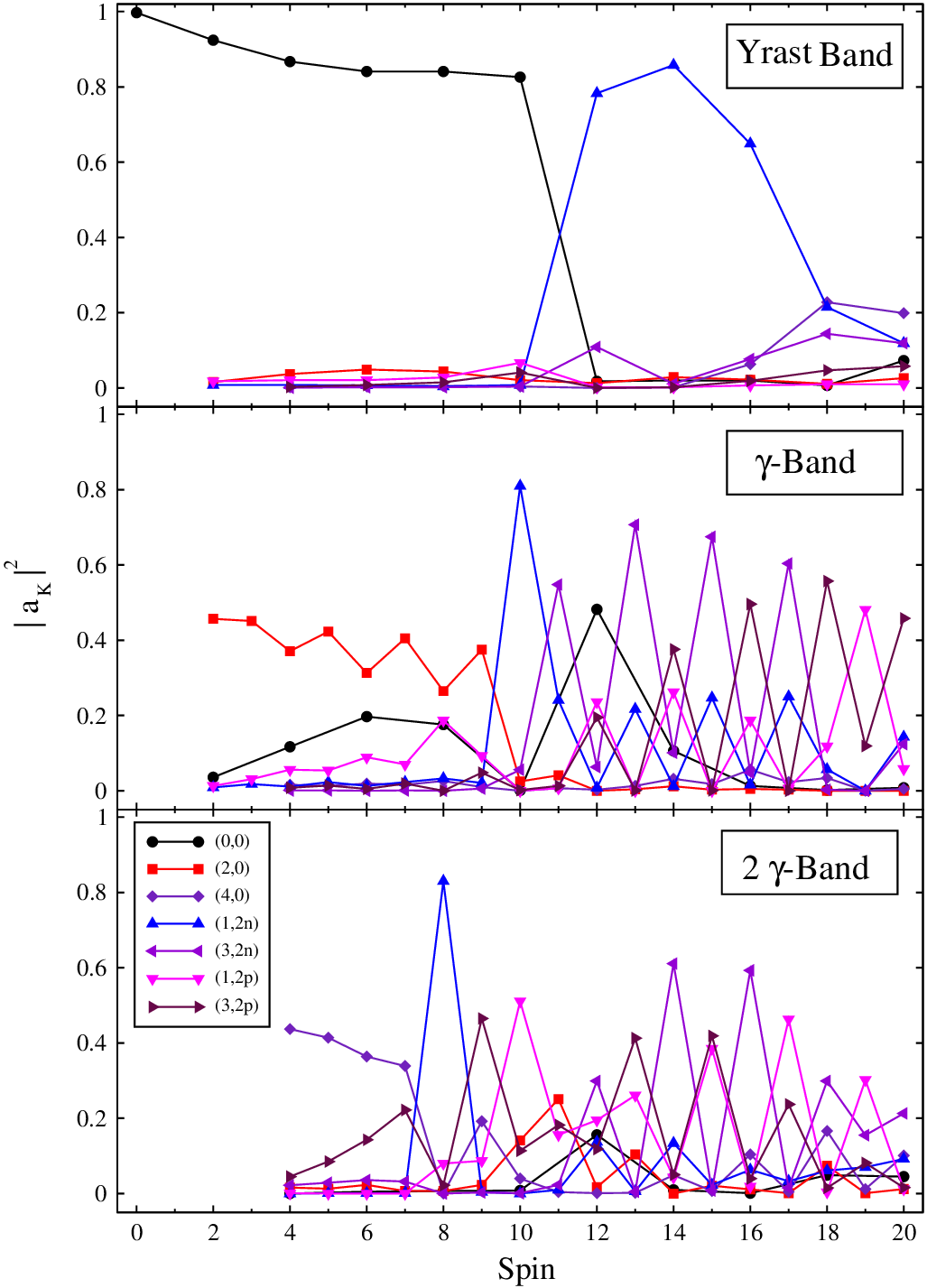}
\caption{ Wavefunction decomposition for $^{156}$Er. $a_K$ denotes the
amplitude of the wavefuction in terms of the projected basis states.
}

\label{figure.7}
\end{center}
\end{figure*}

\begin{figure*}
\begin{center}
\includegraphics[width=6.5in]{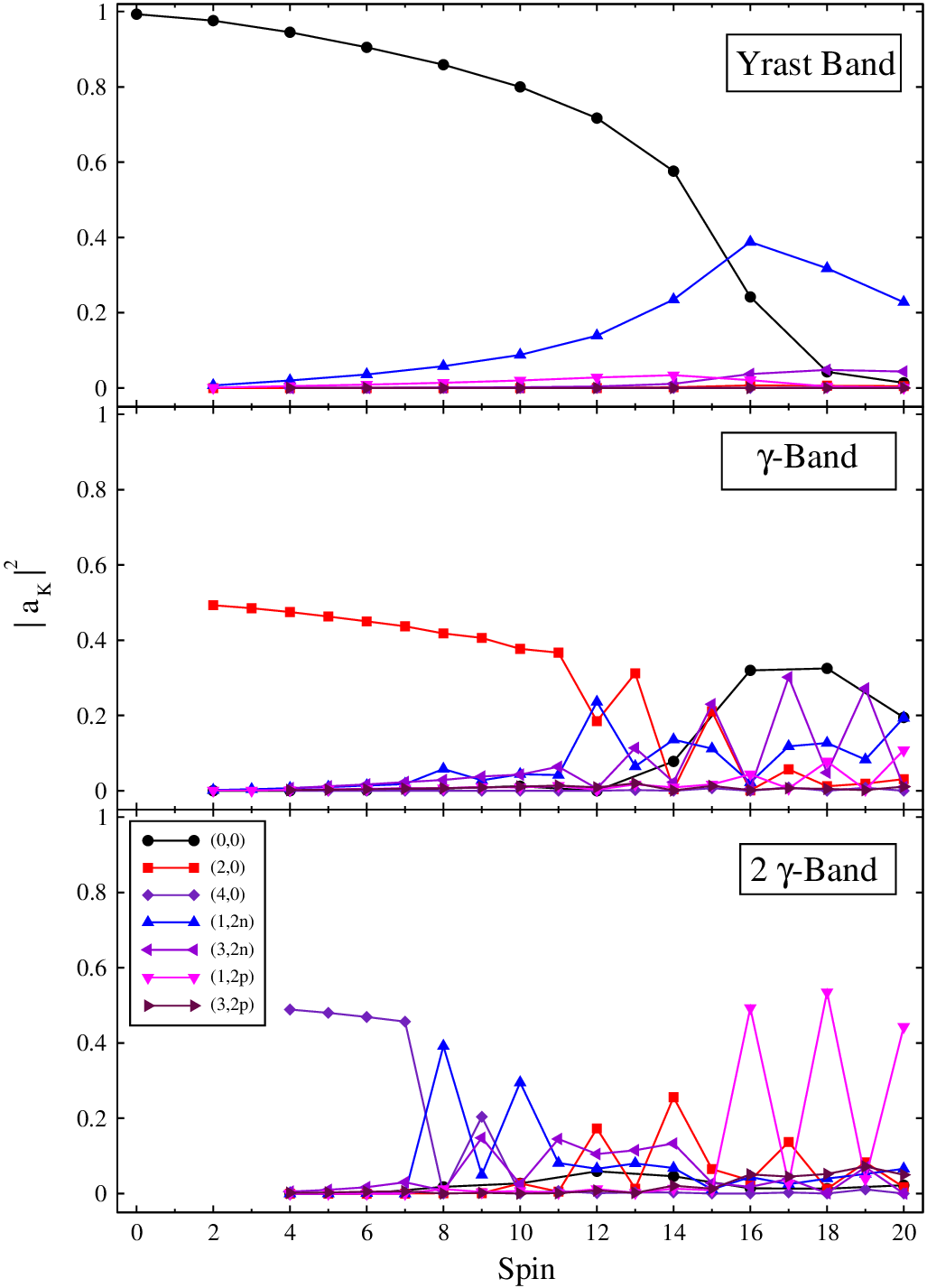}
\caption{ Wavefunction decomposition for $^{164}$Er. $a_K$ denotes the
amplitude of the wavefuction in terms of the projected basis states.}

\label{figure.8}
\end{center}
\end{figure*}

\begin{figure*}
\begin{center}
\includegraphics[width=6.5in]{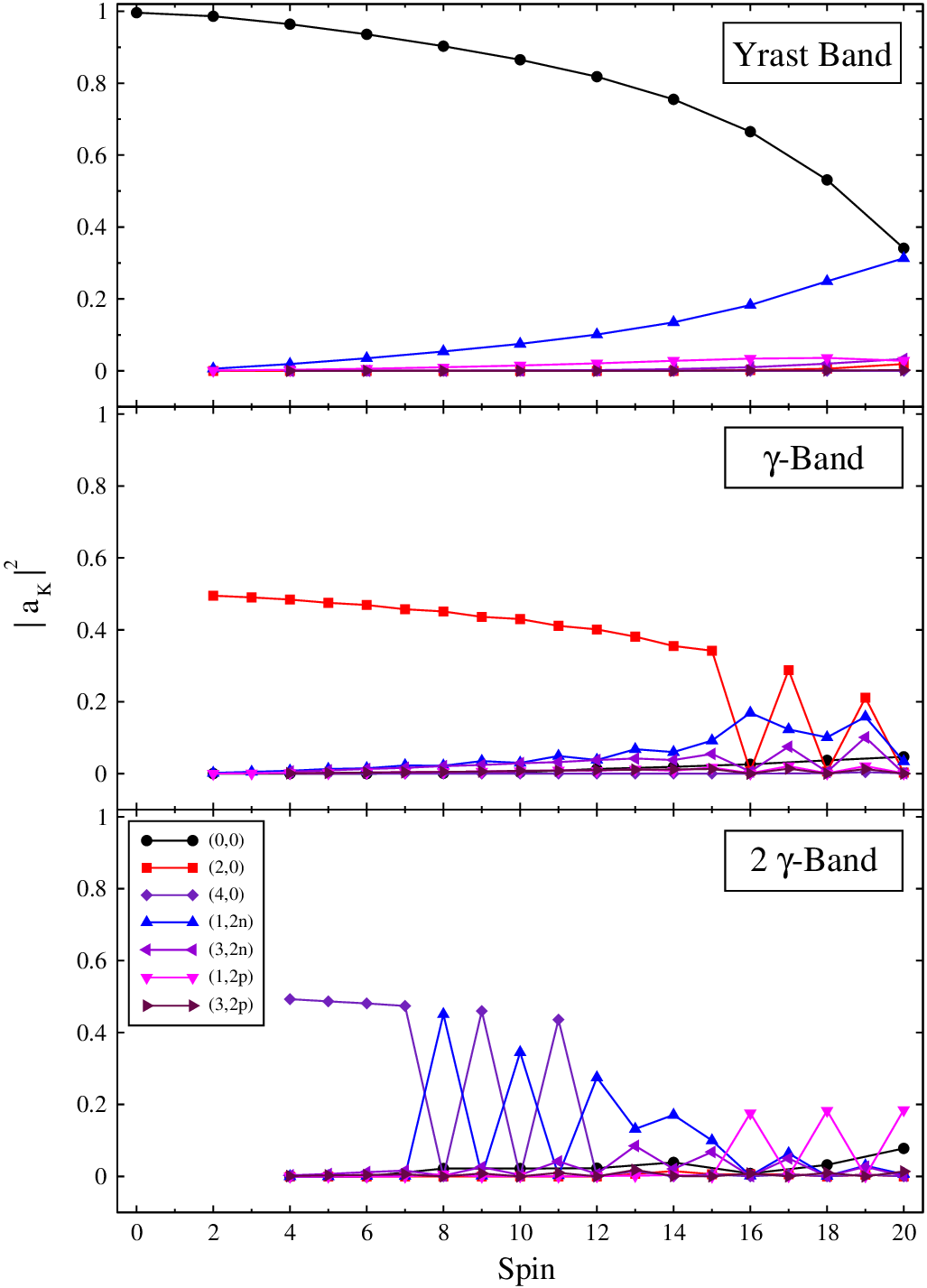}
\caption{ Wavefunction decomposition for $^{170}$Er. $a_K$ denotes the
amplitude of the wavefuction in terms of the projected basis states.}

\label{figure.9}
\end{center}
\end{figure*}

\end{document}